\documentstyle[12pt]{article}
\newcommand{\EQ}{\begin{equation}}
\newcommand{\EN}{\end{equation}}
\newcommand{\bea}{\begin{eqnarray}}
\newcommand{\eea}{\end{eqnarray}}
\newcommand{\hs}{\hspace{0.1cm}}

\newcommand{\th}{\theta}
\newcommand{\var}{\varepsilon}

\newcommand{\goto}{\rightarrow}

\begin{document}
\setcounter{page}{0}
\topmargin 0pt
\oddsidemargin 5mm
\renewcommand{\thefootnote}{\arabic{footnote}}
\newpage
\setcounter{page}{0}
\begin{titlepage}
\begin{flushright}
LPTHE/98-51
\end{flushright}
\vspace{0.5cm}
\begin{center}
{\large {\bf Off-critical correlations in the Ashkin-Teller model}}\\
\vspace{1.8cm}
{\large G. Delfino} \\
\vspace{0.5cm}
{\em Laboratoire de Physique Th\'eorique et Hautes Energies}\\
{\em Universit\'e Pierre et Marie Curie, Tour 16 $1^{er}$ \'etage, 4 place 
Jussieu}\\
{\em 75252 Paris cedex 05, France}\\
{\em E-mail: aldo@lpthe.jussieu.fr}\\
\end{center}
\vspace{1.2cm}

\renewcommand{\thefootnote}{\arabic{footnote}}
\setcounter{footnote}{0}

\begin{abstract}
\noindent
We use the exact scattering description of the scaling Ashkin-Teller model
in two dimensions to compute the two-particle form factors of the relevant 
operators. These provide an approximation for the correlation functions 
whose accuracy is tested against exact sum rules.
\end{abstract}

\vspace{.3cm}

\end{titlepage}

\newpage
\noindent
{\bf 1.} The computation of correlation functions is a central and notoriously 
difficult
problem of quantum field theory and statistical mechanics. The integrable 
models of two-dimensional quantum field theory provide an unique framework 
where such a problem can be studied in a non-perturbative way. Although even
in this context the determination of exact analytic expressions for the 
correlations functions remains a challenging task, it is remarkable
that an extremely accurate evaluation of the correlators is nowdays possible 
and involves a relatively small amount of technical work. The method is quite 
general 
and relies on the solution of the integrable model in terms of an exact 
scattering theory \cite{ZZ}. The natural way to express correlators is then 
as spectral 
series over a complete set of intermediate asymptotic states. The computation
of the matrix elements of the operators (form factors) entering the spectral
series is the crucial intermediate step needed to make contact with the 
space of local operators starting from the purely on-shell solution of the 
model \cite{Karowski,Smirnov}. From the physical point of view, a particularly 
interesting point in the problem of the computation of form factors lies 
in the identification of the solution of the form factor equations 
corresponding to a specific operator. In Refs.\,\cite{immf,DS,DSC}, a set of 
operator-dependent constraints for the form factors was identified which 
allows the solution of this {\em identification problem}. The need for
approximation in this program for the computation of correlators arises only
from our present inability to 
resum exactly the spectral series, what forces us to rely on partial sums.
 
In this note we briefly illustrate how this set of ideas applies to the scaling
limit of the Ashkin-Teller (AT) model. The richness of the spectrum
of relevant operators, the presence of different locality sectors and the 
relation with the Sine-Gordon model are among the features which make this 
example interesting from the theoretical point of view. 

The two-dimensional AT model \cite{AT} can be formulated in terms of two 
planar Ising models coupled by a local four-spin interaction. The lattice 
Hamiltonian reads
\EQ
H_{AT}=\sum_{(m,n)}[J(\sigma_1^m\sigma_1^n+\sigma_2^m\sigma_2^n)+
K\sigma_1^m\sigma_1^n\sigma_2^m\sigma_2^n]\,\,,
\label{lattice}
\EN
where $\sigma^m_{1,2}=\pm 1$ are the two Ising spins at the site $m$, the
sum is over nearest neighbours and the same coupling $J$ has been chosen 
for the Ising models (isotropic case). The model is known to exhibit a 
line of second order phase transition in the space of the lattice couplings
$J$ and $K$ along which the critical exponents vary continously 
\cite{Kadanoff,KB}. In this note we will consider the scaling region 
around this critical line. In such a region the correlation length is much 
larger than the lattice spacing and the model admits a continous description
in terms of the Hamiltonian
\EQ
{\cal H}_{AT}={\cal H}^0_1+{\cal H}^0_2+\tau\,\int d^2x\,(\var_1(x)+\var_2(x))+
\rho\,\int d^2x\,\var_1(x)\var_2(x)\,\,,
\label{continous}
\EN
where ${\cal H}^0_i$ and $\var_i(x)$ denote the fixed point Hamiltonian 
and the energy density operator of the $i$-th Ising model, respectively.
The meaning of the last expression is that the scaling limit of the AT model
can be regarded as a conformal field theory with twice the central charge of
the Ising model (i.e. $c=1/2+1/2=1$) perturbed by the two operators 
${\cal E}\equiv\var_1+\var_2$ and $\var_1\var_2$. 

Consider at first the two critical, non-interacting Ising models, namely 
the Hamiltonian (\ref{continous}) with $\tau=\rho=0$.
The operator $\var_1\var_2$ has twice the scaling dimension of the Ising 
energy density, namely $x_{\var_1\var_2}=2$. Then it is marginal and taking
$\rho\neq 0$ in (\ref{continous}) does not spoil criticality. This means 
that the coupling
$\rho$ parameterises the critical line of the AT model. A critical line 
with central charge $c=1$ can be described in terms of a free massless boson
$\varphi$ (Gaussian model). The correspondence between the critical AT model 
and the Gaussian model was established in Ref.\,\cite{KB} (see also 
\cite{bible}) where, in 
particular, a series of operator identifications was obtained. 
These identifications are summarised in the first two columns of the Table.

We denote by $\sigma_i(x)$ and $\mu_i(x)$ ($i=1,2$) the order (spin) and 
disorder operators of the two Ising models. It is a characteristic feature 
of the AT model that these operators preserve along the whole critical line 
the basic properties they have in the pure Ising model \cite{Kadanoff}. In
particular, they retain the same scaling dimension $x_\sigma=x_\mu=1/8$, and
produce the neutral fermions $\psi_i$ under operator product expansion:
$\sigma_i\times\mu_i\sim\psi_i$. 

The parameter $\beta$, equivalent to $\rho$, identifies the different 
points along the critical line in the Gaussian model. 
The field $\tilde{\varphi}$, dual to $\varphi$, is defined by 
$i\partial_\alpha\tilde{\varphi}=\var_{\alpha\beta}\partial_\beta\varphi$,
with $\alpha=1,2$ labelling the two directions on the plane.
Within the normalisation of the bosonic field $\varphi$ we adopt in this paper,
the vertex operators $e^{i\alpha\varphi}$ and 
$e^{i\alpha\tilde{\varphi}}$ have scaling dimension $\alpha^2/4\pi$, what 
completes the third column of the Table. The point $\rho=0$ at which the two
Ising models are decoupled corresponds to $\beta=\sqrt{4\pi}$. It was argued 
in \cite{KB} that the AT critical line corresponds to the range 
$\sqrt{2\pi}\leq\beta\leq \sqrt{8\pi}$.

When $\tau\neq 0$ in the Hamiltonian (\ref{continous}) the system is moved
away from criticality and develops a finite correlation length.
In view of the identification ${\cal E}\sim\cos\beta\varphi$, we see that the
bosonic equivalent of the resulting massive theory is the Sine-Gordon model
defined by the action
\EQ
{\cal A}_{SG}=\int d^2x\,(\frac{1}{2}\partial_\mu\varphi\partial^\mu\varphi+
\mu\cos\beta\varphi)\,\,,
\label{sg}
\EN
where $\mu$ is a dimensional parameter proportional to $\tau$.
The equivalence of the models (\ref{continous}) and (\ref{sg}) (for 
the range of $\beta$ specified above) immediately enables us to extend to 
the scaling AT model the two basic results known for Sine-Gordon, i.e. its
integrability and its exact scattering description \cite{ZZ}.

The exact particle spectrum of the Sine-Gordon model contains two elementary
excitations of mass $m$, the soliton $A_+$ and the antisoliton $A_-$. 
If from the bosonic point of view they are topological excitations, they 
can also be interpreted as the particles associated to the charged elementary
fermion of the equivalent massive Thirring model \cite{Coleman}. If we write 
$A_\pm=A_1\pm iA_2$, then the neutral excitations $A_1$ and $A_2$ are 
naturally associated to the neutral fermions $\psi_1$ and $\psi_2$ of the 
AT model.

Due to integrability, the scattering of the particles $A_1$ and $A_2$ is 
completely elastic and factorised. It is entirely specified by the 
Faddev-Zamolodchikov algebra\footnote{The on-shell energy and momentum of the
particles are parameterised as $(p^0,p^1)=(m\cosh\th,m\sinh\th)$.}
\EQ
A_i(\th_1)A_j(\th_2)=\sum_{k,l=1,2}\sigma_{ij}^{kl}(\th_1-\th_2)
A_l(\th_2)A_k(\th_1)\,\,.
\label{FZ}
\EN
The non-zero scattering amplitudes are given by \cite{ZZ}
\bea
2\sigma_{11}^{11}(\th)=2\sigma_{22}^{22}(\th)=S(\th)+S_+(\th)\,\,,\nonumber\\
2\sigma_{11}^{22}(\th)=2\sigma_{22}^{11}(\th)=S_+(\th)-S(\th)\,\,,\nonumber\\
2\sigma_{12}^{12}(\th)=2\sigma_{21}^{21}(\th)=S(\th)+S_-(\th)\,\,,\nonumber\\
2\sigma_{12}^{21}(\th)=2\sigma_{21}^{12}(\th)=S(\th)-S_-(\th)\,\,,
\eea
where
\EQ
S_+(\th)=-\frac{\sinh\frac{\pi}{2\xi}(\th+i\pi)}
               {\sinh\frac{\pi}{2\xi}(\th-i\pi)}S(\th)\,,\hspace{1cm}
S_-(\th)=-\frac{\cosh\frac{\pi}{2\xi}(\th+i\pi)}
               {\cosh\frac{\pi}{2\xi}(\th-i\pi)}S(\th)\,,
\EN
\EQ
S(\th)=-\exp\left\{-i\int_0^\infty\frac{dx}{x}\frac{\sinh\frac{x}{2}\left(1-
\frac{\xi}{\pi}\right)}{\sinh\frac{x\xi}{2\pi}\cosh\frac{x}{2}}
\sin\frac{\th x}{\pi}\right\}\,\,,
\EN
\EQ
\xi=\frac{\pi\beta^2}{8\pi-\beta^2}\,\,.
\EN
The interaction among the elementary particles is attractive (repulsive) 
below (above) the free fermion point $\beta=\sqrt{4\pi}$. In the attractive 
regime, the elementary excitations form bound states (breathers). 
Our subsequent analysis can be easily extended to include 
breathers. For the sake of simplicity, however, we will not go into the 
details of the breather sector in this note.

{\bf 2.} The basic physical ingredients entering the computation of form 
factors are conveniently illustrated on the example of the 
two-fermion matrix elements\footnote{We denote
by $\Phi$ a generic scalar (under Lorentz transformations) operator.}
\EQ
F^\Phi_{ij}(\th_1-\th_2)=\langle 0|\Phi(0)|A_i(\th_1)A_j(\th_2)\rangle\,,
\hspace{1cm}i,j=1,2\,.
\label{ff}
\EN
They satisfy the unitarity and crossing relations \cite{Karowski}
\bea
F^\Phi_{ij}(\th)=\sum_{k,l=1,2}\sigma_{ij}^{kl}(\th)F^\Phi_{lk}(-\th)\,\,,
\nonumber\\
F^\Phi_{kl}(\th+2i\pi)=e^{2i\pi\gamma_{\Phi,k}}F^\Phi_{lk}(-\th)\,\,,
\label{monodromy}
\eea
where the factor 
$e^{2i\pi\gamma_{\Phi,k}}$ in the crossing equation takes into account 
a possible non-locality among the operator $\Phi$ and the particle $A_k$
\cite{Smirnov,Yurov,msg}. For the operators we are concerned with in the AT
model we just need to recall that $A_i$ is associated to the fermion $\psi_i$.
The latter is non-local with respect to the operators $\sigma_j$ and $\mu_j$
with non-locality index $\gamma_{\sigma_j,i}=\gamma_{\mu_j,i}=\frac{1}{2}
\delta_{i,j}$; $\psi_i$ is instead local with respect to itself and
$\var_j=\bar\psi_j\psi_j$.
For the operators defined as products, $\Phi=\prod_j\Phi_j$, the 
non-locality index reads $\gamma_{\Phi,i}=\sum_j\gamma_{\Phi_j,i}$. These 
considerations provide the fourth column of the Table.

It is obvious from the form of the lattice Hamiltonian (\ref{lattice}) that
the AT model is invariant under the change of sign of $\sigma_1$ and/or
$\sigma_2$ ($Z_2\times Z_2$--symmetry), and under the exchange 
$1\leftrightarrow 2$ 
(exchange symmetry). The fermions $\psi_i$, and then the particles $A_i$, 
are odd under the $Z_2\times Z_2$--symmetry. The behaviour of the different
operators under this symmetry then determines the asymptotic
states on which each operator has non-zero matrix elements. These states are 
given in the last column of the Table (modulo pairs of the type $A_jA_j$).
The symmetry properties of the operators under the exchange symmetry are 
important to simplify the functional system (\ref{monodromy}).

The scalar operators $\Phi$ which are not odd under any of the global 
symmetries of the model may have a nonvanishing vacuum expectation value 
$\langle\Phi\rangle$. If in addition they are non-local with respect to the 
fermions, their two-particle form factors exhibit a pole at $\th=i\pi$ with
residue
\EQ
-i\mbox{Res}_{\th=i\pi}F^\Phi_{jj}(\th)=(1-e^{2i\pi\gamma_{\Phi,j}})
\langle\Phi\rangle\,\,.
\EN
The only additional singularities allowed on the physical strip $0<\th<i\pi$ 
are the poles associated to the bound states. The functional equations
(\ref{monodromy}), together with the required singularity structure, fix
the matrix elements (\ref{ff}) up to polynomials of $\cosh\th$. The 
asymptotic arguments of Ref.\,\cite{immf} can be used to show that, for the
relevant ($x_\Phi<2$) operators listed in the Table, these polynomials are of
degree zero. The non-zero two-fermion matrix elements are then {\em uniquely}
determined to be ($j,l=1,2$)
\bea
&& F^{\mu_l}_{jj}(\th)=\frac{i\pi\langle\mu\rangle}{2\xi\omega(i\pi)}\,
\frac{F_0(\th)}{\sinh\frac{\pi}{2\xi}(\th-i\pi)}
\left[\omega(\th)+(-1)^{l+j}\omega(2i\pi-\th)\right]\,\,,\\
&& F^{\cal E}_{jj}=c_1\,\frac{\cosh\frac{\th}{2}}{\sinh\frac{\pi}{2\xi}
(\th-i\pi)}\,F_0(\th)\,\,,\label{energy}\\
&& F^{\cal C}_{jj}=c_2\,(-1)^j\,F_0(\th)\,\,,\\
&& F^{{\cal E}_+}_{12}=F^{{\cal E}_+}_{21}=c_3\,F_0(\th)\,\,,\\
&& F^{{\cal E}_-}_{12}=-F^{{\cal E}_-}_{21}=c_4\,\frac{\cosh\frac{\th}{2}}
{\cosh\frac{\pi}{2\xi}(\th-i\pi)}\,F_0(\th)\,\,,\\
&& F^{{\cal P}}_{12}=-F^{{\cal P}}_{21}=c_5\,\frac{F_0(\th)}
{\cosh\frac{\pi}{2\xi}(\th-i\pi)}\,\,,\\
&& F^{{\cal P}^*}_{jj}=\frac{i\pi}{\xi}\langle{\cal P}^*\rangle\,
\frac{F_0(\th)}{\sinh\frac{\pi}{2\xi}(\th-i\pi)}\,\,.
\eea
Here the $c_n$ are normalisation constants, $\langle\mu\rangle\equiv
\langle\mu_1\rangle=\langle\mu_2\rangle$, and the functions
\EQ
\omega(\th)=\exp\left\{2\int_0^\infty\frac{dx}{x}\,
\frac{\sinh\left(1-\frac{\xi}{\pi}\right)x}
{\sinh\frac{x\xi}{\pi}}\,\frac{\sin^2\frac{\th x}{2\pi}}{\sinh 2x}\right\}\,\,,
\EN
\EQ
F_0(\th)=-i\sinh\frac{\th}{2}\,
\exp\left\{\int_0^\infty\frac{dx}{x}\,\frac{\sinh\frac{x}{2}\left(1-
\frac{\xi}{\pi}\right)}{\sinh\frac{x\xi}{2\pi}\,\cosh\frac{x}{2}}\,
\frac{\sin^2\frac{(i\pi-\th)x}{2\pi}}{\sinh x}\right\}\,\,
\EN
satisfy the equations
\EQ
\omega(\th)=\omega(-\th)\,\,,\hspace{1cm}
\omega(\th+2i\pi)=-\frac{\sinh\frac{\pi}{2\xi}(\th+i\pi)}
{\sinh\frac{\pi}{2\xi}(\th-i\pi)}\omega(\th-2i\pi)\,\,,
\EN
\EQ
F_0(\th)=S(\th)F_0(-\th)\,\,,\hspace{1cm}F_0(\th+2i\pi)=F_0(-\th)\,\,.
\EN
For large values of $|\th|$ they behave as 
$\omega(\th)\sim\exp\left[\left(\frac{\pi}{\xi}-1\right)\frac{|\th|}{4}
\right]$ and $F_0(\th)\sim\exp\left[\left(\frac{\pi}{\xi}+1\right)
\frac{|\th|}{4}\right]$.

The form factors of the operators $e^{i\alpha\varphi}$ are known for the
Sine-Gordon model \cite{Karowski,Smirnov,Lukyanov}. Of course,
the two-fermion matrix elements for the operators ${\cal E}$, ${\cal E}_-$, 
${\cal P}$ and ${\cal P}^*$ determined here coincide with those of the 
corresponding bosonic operators in Sine-Gordon.

The notation we use in the first column of the Table for the AT operators 
refers to the high temperature phase of the model. 
The high and low temperature phases, however, are mapped into one another
by the duality transformations $\sigma_i\leftrightarrow\mu_i$,
$\psi_i\leftrightarrow\bar{\psi}_i$. In particular, this implies that 
$\sigma_i$ and $\mu_i$ represent the same operator in the two different
phases. In the form factor approach their relative normalisation is fixed
by the factorisation condition \cite{DSC}
\EQ
\lim_{|\th|\rightarrow\infty}F^{\mu_i}_{ii}(\th)=\frac{\left(F^{\sigma_i}_i
\right)^2}
{\langle\mu\rangle}\,\,,
\EN
wich determines the one-particle matrix element
$F^{\sigma_i}_i=\langle 0|\sigma_i(0)|A_i\rangle$.

Although the generalisation of the functional system (\ref{monodromy}) to the
matrix elements with more than two fermions in the asymptotic state is 
straightforward, its solution poses a non-trivial mathematical problem
\cite{Smirnov}. For the reasons to be explained in a moment, we do not need 
to go into this technical part here.

{\bf 3.} In general, the knowledge of the $n$-particle form factors 
$\langle 0|\Phi(0)|n\rangle$ allows expressing the correlation functions 
in the form of the spectral series
\EQ
\langle\Phi_1(x)\Phi_2(0)\rangle=\sum_{n=0}^\infty\langle 0|\Phi_1(0)|n\rangle
\langle n|\Phi_2(0)|0\rangle e^{-E_n|x|}\,\,,
\label{corr}
\EN
where $E_n$ denotes the total energy of the $n$-particle asymptotic state. 
Although the exact resummation of the spectral series is beyond the present 
technical possibilities, it is known that partial sums provide very good 
approximations (see e.g. Ref.\,\cite{DC} for a more detailed discussion and
additional references on this 
subject). Remarkably, this is already the case for the ``two-particle 
approximation'', which corresponds to neglecting the contribution of all the 
states with $n>2$ in the spectral series (\ref{corr}). The fermionic matrix
elements computed in this note are what is needed to implement the 
two-particle
approximation in the breatherless region $4\pi\leq\beta^2\leq 8\pi$. As a 
quantitative check of the convergence of the spectral series, we evaluate 
in this approximation the central charge and the scaling dimension $x_\mu$
through the sum rules \cite{cth,Cardycth,DSC}
\bea
c=\frac{3}{4\pi}\int d^2x\,|x|^2\langle\Theta(x)\Theta(0)\rangle_c\,\,,
\label{cth}\\
x_\mu=-\frac{1}{2\pi\langle\mu\rangle}\int d^2x\,\langle\Theta(x)\mu(0)
\rangle_c\,\,,
\eea
where $\langle\cdots\rangle_c$ denotes connected correlators and 
$\Theta(x)$ is the trace of the stress-energy tensor. The latter is 
proportional to the energy operator ${\cal E}(x)$ and its two-fermion form 
factors are given by (\ref{energy}) with the normalisation fixed 
by\footnote{Our normalisation for the asymptotic
states is defined by $\langle A(\th)|A(\th')\rangle=2\pi\delta(\th-\th')$.}
$F^\Theta_{jj}(i\pi)=2\pi m^2$.
Numerical integration shows that the two-fermion contribution to the central
charge decreases monotonically from $1$ at $\beta^2=4\pi$ to $0.987$ at
$\beta^2=8\pi$. For the two-fermion contribution to $x_\mu$ one finds a 
monotonic decrease from $0.125$ to $0.110$ over the same range. The 
circumstance that
this computation gives the exact answer at $\beta^2=4\pi$ simply follows from
the fact that $\Theta\sim{\cal E}=\bar{\psi}_1\psi_1+\bar{\psi}_2\psi_2$
couples only to the two-particle states at the free fermion point. The fact
that the approximation is more accurate for $c$ than for $x_\mu$ away from the 
free point is due to the 
factor $|x|^2$ in the integral (\ref{cth}) which suppresses the contribution
of the short distances. This is the region where our error localises when we
truncate the large distance expansion (\ref{corr}).

\vspace{1cm}
\noindent
{\bf Acknowledgements.} I thank V.A. Fateev for discussions.

\newpage

\begin{center}

\vspace{3cm}
\begin{tabular}{|c|c|c|c|c|c|c|}\hline
$\Phi$ & Bosonic & $x_\Phi$ & $\gamma_{\Phi,j}$ & 
$Z_2\times Z_2$ & $1\leftrightarrow 2$ & Fermion  \\ 
& form & & & & & sector \\ \hline
$\sigma_i$ & & $\frac{1}{8}$ & $\frac{1}{2}\delta_{i,j}$ & $(-)^i\times
(-)^{i+1}$ & & $A_i$ \\
$\mu_i$ & & $\frac{1}{8}$ & $\frac{1}{2}\delta_{i,j}$ & $+\times +$ & & 
$A_jA_j$ \\
${\cal E}=\var_1+\var_2$ & $\cos\beta\varphi$ & $\frac{\beta^2}{4\pi}$ & 
$0$ & $+\times +$ & + & $A_jA_j$ \\
${\cal C}=\var_1-\var_2$ & $\cos\frac{4\pi}{\beta}\tilde{\varphi}$ & 
$\frac{4\pi}{\beta^2}$ & $0$ & $+\times +$ & $-$ & $A_jA_j$ \\
${\cal E}_+=\bar{\psi}_1\psi_2+\bar{\psi}_2\psi_1$ & 
$\sin\frac{4\pi}{\beta}\tilde{\varphi}$ & 
$\frac{4\pi}{\beta^2}$ & $0$ & $-\times -$ & + & $A_1A_2$ \\
${\cal E}_-=\bar{\psi}_1\psi_2-\bar{\psi}_2\psi_1$ & 
$\sin\beta\varphi$ & $\frac{\beta^2}{4\pi}$ & 
$0$ & $-\times -$ & $-$ & $A_1A_2$ \\
${\cal P}=\sigma_1\sigma_2-\sigma_2\sigma_1$ & 
$\sin\frac{\beta}{2}\varphi$ & $\frac{\beta^2}{16\pi}$ & 
$\frac{1}{2}$ & $-\times -$ & $-$ & $A_1A_2$ \\
${\cal P}^*=\mu_1\mu_2+\mu_2\mu_1$ & $\cos\frac{\beta}{2}\varphi$ & 
$\frac{\beta^2}{16\pi}$ & $\frac{1}{2}$ & $+\times +$ & $+$ & $A_jA_j$ \\
$\sigma_1\mu_2$ & $\cos\frac{2\pi}{\beta}\tilde{\varphi}$ & 
$\frac{\pi}{\beta^2}$ & $\frac{1}{2}$ & $-\times +$ & & $A_1$ \\
$\mu_1\sigma_2$ & $\sin\frac{2\pi}{\beta}\tilde{\varphi}$ & 
$\frac{\pi}{\beta^2}$ & $\frac{1}{2}$ & $+\times -$ & & $A_2$ \\ \hline
\end{tabular}
\end{center}

\begin{center}
{\bf Table}
\end{center}

\end{document}